\documentclass{ws-rv975x65}  

\begin{document}

\setcounter{chapter}{0}

\centerline{MONOPOLES AND PROJECTIVE REPRESENTATIONS: TWO AREAS}
\centerline{OF INFLUENCE OF YANG--MILLS THEORY ON MATHEMATICS}

\bigskip

\markboth{S. Adler}{Monopoles \& Projective Representations: Two Areas of Influence of Yang--Mills Theory...}

\author{Stephen L. Adler}

\address{Institute for Advanced Study\\
Einstein Drive, Princeton, NJ 08540\\
E-mail: adler@ias.edu}

\begin{abstract}
I describe how my involvement with monopoles related to the 
multimonopole existence proof of Taubes, and how my later work on 
quaternionic quantum mechanics led to 
the classification theorem for generalized projective group representations 
of Tao and Millard.  
\end{abstract}

\section{Introduction \label{intro}} 

In this essay I discuss two unrelated subjects with a common theme: the 
influence of Yang--Mills theory on mathematics. In the first part I describe 
how my interest in monopole solutions in the period 1978--1980 led indirectly 
to the completion by Clifford Taubes of his multimonopole existence theorem 
during a visit to the Institute for Advanced Study in the spring of 1980. 
In the second part I shift my focus to a generalized definition of projective 
group representations that I proposed in the context of quaternionic 
quantum mechanics. This inspired a  classification theorem proved 
by Terry Tao and Andrew Millard in 1996, which has useful implications 
for projective representations in standard, complex quantum mechanics.

\section{Multimonopole Solutions \label{multi}}

During the years 1978--1980 I became interested in trying to find a simple, 
semi-classical model for quark confinement. My first attempt, which did 
not succeed but which had useful by-products that I shall describe here,  
was based on the idea of considering the potential between classical 
quark sources in the background of a non-Abelian 't Hooft--Polyakov [\refcite{thooft}] 
Prasad--Sommerfield [\refcite{pra-som}]
monopole or its generalizations, which I conjectured [\refcite{sla-78}] might act 
as a quark-confining ``bag''.
To pursue (and ultimately rule out) this conjecture, I did a number of 
calculations of properties of monopole solutions. The first was a calculation 
of the Green's function for a single Prasad--Sommerfield monopole, by using 
the multi-instanton representation of the monopole and a formalism for  
calculating multi-instanton Green's functions given by Brown et al [\refcite{bccl}].  
This calculation was spread over two papers that I wrote; setting up contour  
integral expressions for the Green's function was done in Appendix A 
of Adler [\refcite{sla-78}], and the final result for the monopole propagator, after 
evaluation 
of the contour integrals and considerable algebraic simplification, was 
given in Appendix A of Adler [\refcite{sla79}].  From the propagator formula, I concluded   
that a single monopole background would not lead to confinement.

Not yet ready to give up on the monopole background idea, I then wrote two  
papers speculating that the Prasad--Sommerfield monopole might 
be a member of a larger class of solutions, in which the point at which the 
monopole Higgs field vanishes is extended to a higher-dimensional region,  
and in particular to a ``string''-like configuration with a line segment 
as a zero set. In the first of these papers [\refcite{sla-small}], I studied small 
deformations around the Prasad--Sommerfield monopole and found several series 
of such deformations.  For normalized deformations I recovered the monopole 
zero modes that had already been obtained by Mottola [\refcite{mott}], but I found that 
``if an axially symmetric extension exists, it cannot be reached by 
integration out along a tangent vector defined by a nonvanishing, 
non-singular small-perturbation mode''.  This work was later extended into 
a complete calculation of the perturbations around the Prasad--Sommerfield 
solution by Akhoury, Jun, and Goldhaber [\refcite{akh}], who also found ``no acceptable 
nontrivial zero energy modes.''  In my second paper [\refcite{sla-global}], 
I employed nonperturbative methods and suggested that despite the negative 
perturbative results, there might still be interesting extensions of the 
Prasad--Sommerfield solution with extended Higgs field zero sets.  

At just around the same time, Erick Weinberg  
wrote a paper [\refcite{ejw}] extending an index theorem of 
Callias [\refcite{callias}] to give a parameter counting theorem for multi-monopole 
solutions.  Weinberg concluded that ``any solution with $n$ units 
of magnetic charge belongs to a $(4n-1)$-parameter family of solutions.  It  
is conjectured that these parameters correspond to the positions and 
relative $U(1)$ orientations of $n$ noninteracting unit monopoles''.  For 
$n=1$, his results agreed with the zero-mode counting implied by Mottola's 
explicit calculation.  Weinberg and I were aware of each other's work, as 
evidenced by correspondence in my file dating from March to June of 1979, 
and references relating to this correspondence in our papers [\refcite{sla-global}] and [\refcite{ejw}].  

My contact with Clifford Taubes was initiated by an April, 1979 letter from 
Arthur Jaffe, after I gave a talk at Harvard while 
Jaffe, as it happened, was visiting Princeton!  In his letter, Jaffe  
noted that I was working on problems similar to those on 
which his students were working, and enclosed a copy of a paper by 
Clifford Taubes.   (This preprint was not filed with Jaffe's letter, so I 
am not sure which of the early Taubes papers listed on the SLAC Spires 
archive that it was.)  
Jaffe's letter initiated 
telephone contacts with Taubes and some correspondence from him.  On Jan. 6, 
1980 Taubes wrote to me that he was making progress in proving the existence 
of multi-monopole Prasad--Sommerfield solutions, and in this letter and a 
second one dated on January 18, 1980 he reported results that were relevant 
to my conjectures on the possibility of deformed monopoles.  His results 
placed significant restrictions on my conjectures; in a letter dated 
Feb. 1, 1980 I wrote to Lochlainn O'Raifeartaigh, who had also 
been interested in axially symmetric monopoles, saying that 
``On thinking some more 
about your paper (O'Raifeartaigh's preprint was unfortunately not 
retained in my files) I 
realized that the enclosed argument by Cliff Taubes is strong evidence 
against $n=2$ monopoles involving a line zero.  What Taubes shows is that 
a finite action solution of the Yang--Mills--Higgs Lagrangian cannot have 
a line zero of arbitrarily great length; hence if $n=2$ monopoles contained 
a line zero joining the monopole centers, the monopole separation would be 
bounded from above.  But this seems unlikely....''   This correspondence 
and the result of Taubes was mentioned at the end of the published version, 
Houston and O'Raifeartaigh [\refcite{houston}].  

As a result of our overlapping interests, I arranged for Taubes to make 
an informal visit, of two or three months, to the Institute 
for Advanced Study during the spring of 1980.  Clifford had expressed 
interest in this, he noted in a recent email, in part 
because Raoul Bott had suggested 
that he visit the Institute to get acquainted with Karen Uhlenbeck, who 
was visiting the IAS that year.  In the course of his visit he met and  
interacted with Uhlenbeck, who, along with Bott, had a major 
impact on his development as a mathematician.  

Taubes began the visit by looking at my conjecture 
of extended zero sets, but after a while told me that he could not 
find an argument for them.  Partly as a result of his work, I was getting 
disillusioned with my own conjecture, so I asked him what was happening 
with his attempted proof of multi-monopole solutions.  Taubes replied that 
he was stuck on that, and not sure whether they existed. I then mentioned 
to him Erick Weinberg's parameter counting result, which strongly suggested 
a space of moduli much like that in the instanton case, where looking at 
deformations correctly suggests the existence and structure of the 
multi-instanton solutions.  To my surprise, Taubes was not aware of Erick's 
result, and knowing it impelled him into action on his multi-monopole proof. 
Within a week or two he had completed a proof,\footnote{Thus, there was a parallel to what happened a year before with respect to solutions of the first order Ginzburg--Landau equations.  In that case Taubes had heard a lecture at Harvard by E. Weinberg on parameter counting for multi-vortex solutions (written up as Weinberg [\refcite{ewein}]) and then went home and came up with his existence proof for multi-vortices (Taubes [\refcite{taubes}]).  The vortex work provided the initial impetus for Taubes' turning to the monopole problem.} and wrote it up on his  
return to Harvard.  In his paper he showed that ``for every 
integer $N\not=0$ there is at least a countably infinite set of solutions 
to the static $SU(2)$ Yang-Mills-Higgs equations in the Prasad--Sommerfield 
limit with monopole number $N$.  The solutions are partially parameterized 
by an infinite sublattice in $S_N(R^3)$, the $N$-fold symmetric product 
of $R^3$ and correspond to noninteracting, distinct monopoles.'' This 
quote is taken from the Abstract of his preprint ``The Existence of 
Multi-Monopole Solutions to the Static, $SU(2)$ Yang-Mills-Higgs Equations 
in the Prasad-Sommerfield Limit'', which was received on the SLAC Spires 
data base in June, 1980, and which carried an acknowledgement on the 
title page noting that ``This work was completed while the author was a guest 
at the Institute for Advanced Studies, Princeton, NJ 08540''.  
His preprint also ended with an Acknowledgment section
noting his conversations with me, with Arthur Jaffe, and 
with Karen Uhlenbeck, as well as the Institute's hospitality.  The proof    
was not published in this form, however, but instead appeared (with  
acknowledgments edited out at some stage) as Chapter IV of the book by 
Arthur Jaffe and 
Clifford Taubes, ``Vortices and Monopoles'' (Birkh\"auser Boston, 1980), 
that was completed soon afterwards, in August of 1980. The multimonopole 
existence proof was a milestone in Taubes' career; in a recent exchange 
of emails 
relating to this essay, Taubes commented on his visit ``to hang out at the 
IAS during the spring of 1980.  It profoundly affected my subsequent 
career...'' He went on to further investigations of monopole solutions, that 
lead him to studies of 4-manifold theory which have had a great impact on   
mathematics. 

O'Raifeartaigh, who had been following the monopole work at a distance, 
invited me during the spring of 1980 to come to Dublin that summer to lecture 
on my papers. However, since Taubes had much more interesting results 
I suggested to Lochlainn that he ask Clifford instead, and Taubes did go to  
Dublin to lecture. After Clifford's visit, I redirected my search for   
semiclassical confinement models to a study of nonlinear dielectric models 
by analytic and numerical methods, in collaboration  with Tsvi Piran; these 
models do give 
an interesting class of confining theories.  Based on the observation that  
the Yang--Mills action is multiquadratic, Piran and I also applied 
the same numerical 
relaxation methods to give an efficient method for the 
computation of axially symmetric multimonopole 
solutions.  This was described in our Reviews of Modern Physics article [\refcite{sla84}] 
that marked the completion of the research program on confining models, and     
as a by-product, on monopoles.\footnote{Our numerical calculations in the 2-monopole case served mainly to illustrate the computer methods, since by then exact analytic 2-monopole solutions had appeared; see Forgacs, Horvath, and Palla [\refcite{fhp81}] and Ward [\refcite{fhp81}].}

\section{Projective Group Representations \label{pgr}}

Given two group elements $b,a$ with product $ba$, a unitary operator 
representation 
$U_b$ in a Hilbert space is defined by $U_bU_a=U_{ba}$.  A more general 
type of representation, called a ray or projective representation, is 
relevant to describing the symmetries of quantum mechanical systems.  In 
his famous paper on unitary ray representations of Lie groups, 
Bargmann [\refcite{barg}] defines a projective representation as one obeying 
\begin{equation}
U_bU_a=U_{ba}\omega(b,a) ~~~,
\label{ask}
\end{equation}  
with $\omega(b,a)$ a complex phase.  

This 
definition is familiar, and seems obvious, until one asks the following 
question: Eq.~(\ref{ask}) is assumed to hold as an operator identity when acting  
on {\it all} states in Hilbert space.  However, we know that it suffices  
to specify the action of an operator on {\it one} complete set of states 
in Hilbert space to specify the operator completely.  Hence why does one 
not start instead from the definition 
\begin{equation}
U_bU_a|f\rangle=U_{ba}|f\rangle \omega(f;b,a)~~~, 
\label{box}
\end{equation} 
with $\{|f\rangle\}$  one complete set of states, as defining a projective  
representation in Hilbert space?  Let us call Bargmann's definition of 
Eq.~(\ref{ask}) a ``strong'' projective representation, and the definition 
of Eq.~(\ref{box}) a ``weak'' projective representation.  Then  the question 
becomes that of finding the relation between weak and strong projective 
representations.  

Although I have formulated this question here in complex Hilbert space, it 
arose and was solved in the context of quaternionic Hilbert space, where 
the phases $\omega(f;b,a)$ are quaternions, which obey a non-Abelian 
or Yang--Mills  group multiplication law isomorphic to $SO(3) \simeq SU(2)$.   
The ``strong'' definition of Eq.~(\ref{ask}) was adopted for the quaternionic 
case by Emch [\refcite{emch}], but in my book on quaternionic quantum mechanics [\refcite{sla95}] I introduced the ``weak'' definition of Eq.~(\ref{box}) 
in order for 
quaternionic projective representations to include  embeddings of nontrivial 
complex projective representations into quaternionic Hilbert space; the 
state dependence of the phase is necessary because even a complex phase 
$\omega$ does not commute with general quaternionic rephasings of the state 
vector $|f\rangle$.  I noted in my book that Eq.~(\ref{box}) can be extended to 
an operator relation by defining 
\begin{equation}
\Omega(b,a) =\sum_f|f\rangle \omega(f;b,a)\langle f| ~~~,
\label{car}
\end{equation}
so that Eq.~(\ref{box}) takes the form
\begin{equation} 
U_bU_a=U_{ba}\Omega(b,a)~~~,
\label{dot}
\end{equation}
which gives the general operator form taken by projective representations 
in quaternionic quantum mechanics.  I also introduced two specializations 
of this definition, motivated by the commutativity properties of the phase 
factor in complex projective representations.
I defined [\refcite{sla95}]  
a {\it multicentral} 
projective representation as one for which\footnote{In Eq.~(4.51a)   
of [\refcite{sla95}], $U_{ab}$ should read $U_{ba}$, that is, the only conditions are 
those already implied by Eq.~(\ref{elf}).}
\begin{equation}
[\Omega(b,a),U_a]= [\Omega(b,a),U_b]= 0 ~~~ 
\label{elf}
\end{equation}
for all pairs $b,a$, and I defined a {\it central} projective representation 
as one for which 
\begin{equation}
[\Omega(b,a),U_c]=0~~~
\label{fax}
\end{equation}
for all triples $a,b,c$.  

Subsequent to the completion of my book, I read Weinberg's first volume 
on quantum field theory [\refcite{sw}] and realized, from his 
discussion in Sec. 2.7 of the associativity condition for complex 
projective representations, that there must be an analogous associativity 
condition for 
weak quaternionic projective representations.  I worked this out [\refcite{adler96}], and 
showed that it takes the operator form 
\begin{equation}
U_a^{-1}\Omega(c,b)U_a=\Omega(cb,a)^{-1}\Omega(c,ba)\Omega(b,a)~~~,
\label{gem}
\end{equation}
which by the definition of Eq.~(\ref{car}) shows that $U_a^{-1}\Omega(c,b)U_a$ 
is diagonal in the basis $\{|f\rangle\}$, with the spectral representation  
\begin{equation}
U_a^{-1}\Omega(c,b)U_a=\sum_f |f\rangle \overline{\omega(f; cb,a)}
\omega(f;c,ba)\omega(f;b,a)  \langle f|~~~.
\label{hip}
\end{equation}
On the basis of some further identities, I also raised the question [\refcite{adler96}] of 
whether one can construct a multicentral representation that is not central, 
or whether a multicentral representation is always central.  

Subsequently, I discussed the issues of quaternionic projective 
representations with Andrew Millard, who was my thesis student in the   
mid-1990's.  He explained them to his roommate Terry Tao, a mathematics 
graduate student working for Elias Stein, and at my next conference with 
Andrew, Tao came along and presented the outline of a beautiful theorem  
that resolved the issues.  This was written up as a paper of Tao and Millard [\refcite{ttao}], 
and consists of two parts.  The first part is an algebraic analysis based on 
Eq.~(\ref{hip}), which leads to the following theorem  

\bigskip

{\bf Structure Theorem:} {\it Let $U$ be an irreducible projective 
representation of a connected Lie group $G$.  There then exists a reraying 
of the basis $|f\rangle$ under which one of the following three possibilities 
must hold.}

\begin{enumerate}
\item  $U$ is a real projective representation.  That is, $\omega(f;b,a)
=\omega(b,a)$ is independent of $|f\rangle$ and is equal to $\pm 1$ for 
each $b$ and $a$. \hfill\break  
\item  $U$ is the extension of a complex projective representation.  
That is, the matrix elements $\langle f|U_a|f^{\prime}\rangle$ are complex 
and $\omega(f;b,a)=\omega(b,a)$ is independent of $|f\rangle$ and is 
a complex phase. \hfill\break 
\item $U$ is the tensor product of a real projective representation and 
a quaternionic phase.  That is, there exists a decomposition 
$U_a=U_a^{\cal B}\sum_f|f\rangle \sigma_a \langle f|$, where the unitary 
operator $U_a^{\cal B}$ has real matrix elements, $\sigma_a$ is a 
quaternionic phase, and $U_{ba}^{\cal B}=\pm U_b^{\cal B}U_a^{\cal B}$ for 
all $b$ and $a$. 
\end{enumerate}

{}From the point of view of the Structure Theorem, case (1) corresponds to 
the only possibility allowed by the strong definition of quaternionic 
projective representations, as demonstrated earlier by Emch [\refcite{emch}], 
while case (2) corresponds to an embedding of a complex projective 
representation in quaternionic Hilbert space, the consideration of which was 
my motivation for proposing the weak definition. 
Specializing the Structure Theorem to a complex Hilbert space, where  
case (3) cannot be realized, we see that in complex Hilbert 
space the weak projective representation defined in Eq.~(\ref{box}) 
{\it implies} 
the strong projective representation defined in Eq.~(\ref{ask}); hence no generality
is lost by starting from the strong definition, as in Bargmann's paper.  

The second part of the Tao--Millard paper was a proof, by real analysis 
methods, of a Corollary to the structure theorem, stating 

\begin{corollary} Any multicentral projective representation of a 
connected Lie group is central.
\end{corollary}

This thus solved the question of the relation of centrality to 
multicentrality that I raised in my paper [\refcite{adler96}].

Subsequent to this work, I had an exchange with Gerard Emch in the Journal 
of Mathematical Physics [\refcite{gg,sla-resp}] debating the merits of the strong and 
weak definitions.  After a visit to Gainesville where we reconciled 
differing notations, we wrote a joint paper [\refcite{adler97}] clarifying the situation, 
and reexpressing the strong and weak definitions of Eqs.~(\ref{ask}) and (\ref{box}) 
in the language and notation often employed in 
mathematical discussions of projective group representations.  

\section*{Acknowledgments}

This work was supported in part by the Department of Energy under
Grant \#DE--FG02--90ER40542.  I wish to thank Clifford Taubes for several 
emails that helped reconstruct details of his visit and its impact on his
work.

\end{document}